\documentclass{aastex701}

\usepackage{amsmath}
\usepackage{array}
\usepackage{capt-of}
\usepackage{caption}

\begin{document}

\title{\large{On the importance of medium-resolution continuum spectroscopy for JWST's extragalactic legacy}\vspace{0.5cm}}

\author[0000-0002-1482-5818]{Adam C. Carnall}
\affiliation{Institute for Astronomy, University of Edinburgh, Royal Observatory, Edinburgh, EH9 3HJ, UK}
\email{adamc@roe.ac.uk}

\author[orcid=0000-0003-0486-5178]{Ho-Hin Leung}
\affiliation{Institute for Astronomy, University of Edinburgh, Royal Observatory, Edinburgh, EH9 3HJ, UK}
\email{hleung2@roe.ac.uk}

\author[0000-0001-5642-752X]{Struan D. Stevenson}
\affiliation{Institute for Astronomy, University of Edinburgh, Royal Observatory, Edinburgh, EH9 3HJ, UK}
\email{struan.stevenson@ed.ac.uk}

\author[0000-0001-8728-2984]{Elizabeth Taylor}
\affiliation{Institute for Astronomy, University of Edinburgh, Royal Observatory, Edinburgh, EH9 3HJ, UK}
\email{etaylo2@ed.ac.uk}

\author[0000-0002-5258-8761]{Abdurro'uf}\email{fnuabdur@iu.edu}
\affiliation{Department of Astronomy, Indiana University, 727 East Third Street, Bloomington, IN 47405, USA}

\author[0009-0006-2644-6424]{Stella Antonogiannaki}
\affiliation{Institute for Astronomy, University of Edinburgh, Royal Observatory, Edinburgh, EH9 3HJ, UK}
\email{M.S.Antonogiannaki@sms.ed.ac.uk}

\author[0000-0002-2644-3518]{Karla Z. Arellano-C\'{o}rdova}
\affiliation{Centro de Estudios de F\'isica del Cosmos de Arag\'on (CEFCA), Plaza San Juan 1, 44001 Teruel, Spain}
\email{karellano@cefca.es}

\author[0000-0003-0215-1104]{William M. Baker}
\affiliation{DARK, Niels Bohr Institute, University of Copenhagen, Jagtvej 155A, DK-2200 Copenhagen, Denmark}
\email{william.baker@nbi.ku.dk}

\author[0000-0001-6813-875X]{Guillermo Barro}
\affiliation{University of the Pacific, Stockton, CA 90340 USA}
\email{gbarro@pacific.edu}

\author[0000-0001-8104-9751]{Arghyadeep Basu}
\affiliation{Univ Lyon, Univ Lyon1, Ens de Lyon, CNRS, CRAL UMR5574, F-69230, Saint-Genis-Laval, France}
\email{arghyadeep.basu@univ-lyon1.fr}

\author[0000-0003-0629-8074]{Ryan Begley}
\affiliation{Armagh Observatory and Planetarium, College Hill, Armagh, BT61 9DG, N. Ireland, UK}
\affiliation{Institute for Astronomy, University of Edinburgh, Royal Observatory, Edinburgh, EH9 3HJ, UK}
\email{ryan.begley@armagh.ac.uk}

\author[0000-0002-5615-6018]{Sirio Belli}
\affiliation{Department of Physics and Astronomy, University of Bologna, 40129 Bologna, Italy}
\email{sirio.belli@unibo.it}

\author[0000-0002-4153-053X]{Danielle A. Berg}
\affiliation{Department of Astronomy, University of Texas at Austin, 2515 Speedway, Austin, TX 78712, USA}
\affiliation{Cosmic Frontier Center, The University of Texas at Austin, Austin, TX 78712, USA}
\email{daberg@austin.utexas.edu}

\author[0000-0002-9861-4515]{Aliza G. Beverage}
\affiliation{Observatories of the Carnegie Institution for Science, 813 Santa Barbara Street, Pasadena, CA 91101, USA}
\email{abeverage@carnegiescience.edu}

\author[0000-0001-5063-8254]{Rachel Bezanson}
\affiliation{Department of Physics and Astronomy and PITT PACC, University of Pittsburgh, Pittsburgh, PA 15260, USA}
\email{rachel.bezanson@pitt.edu}

\author[0000-0003-4594-6943]{Souradeep Bhattacharya}
\affiliation{Centre for Astrophysics Research, Department of Physics, Astronomy and Mathematics, University of Hertfordshire, Hatfield AL10 9AB, UK}
\email{s.bhattacharya3@herts.ac.uk}

\author[0009-0008-3775-5112]{Cecilia Bondestam}
\affiliation{Institute for Astronomy, University of Edinburgh, Royal Observatory, Edinburgh, EH9 3HJ, UK}
\email{c.bondestam@ed.ac.uk}

\author[0000-0002-8651-9879]{Andrew J.\ Bunker}
\affiliation{Department of Physics, University of Oxford, Denys Wilkinson Building, Keble Road, Oxford, OX1 3RH, UK}
\email{andy.bunker@physics.ox.ac.uk}

\author[0000-0001-8174-317X]{Emilie Burnham}
\affiliation{Department of Astronomy \& Astrophysics, The Pennsylvania State University, University Park, PA 16802, USA}
\email{efb5552@psu.edu}

\author[0000-0002-0450-7306]{Alex J. Cameron}
\affiliation{Cosmic Dawn Center (DAWN), Copenhagen, Denmark}
\affiliation{Niels Bohr Institute, University of Copenhagen, Jagtvej 128, DK-2200, Copenhagen, Denmark}
\email{alex.cameron@nbi.ku.dk}

\author[orcid=0000-0003-2251-9164]{Chloe M. Cheng}
\affiliation{David A. Dunlap Department of Astronomy and Astrophysics, University of Toronto, Toronto, ON M5S 3H4, Canada}
\email{chloe.cheng@astro.utoronto.ca}  

\author[0000-0001-8551-071X]{Yingjie Cheng}
\affiliation{Department of Astronomy, University of Washington, Seattle, WA 98195, USA}
\email{yingjiec@uw.edu}

\author[0000-0003-1371-6019]{M. C. Cooper}
\affiliation{Department of Physics \& Astronomy, University of California Irvine, 4129 Reines Hall, Irvine, CA 92697, USA}
\email{cooper@uci.edu}

\author[0000-0002-1803-794X]{Isa G. Cox}
\email{isacoxastro@gmail.com}
\affiliation{Laboratory for Multiwavelength Astrophysics, School of Physics and Astronomy, Rochester Institute of Technology, 84 Lomb Memorial Drive, Rochester, NY 14623, USA}

\author[0000-0002-3736-476X]{Fergus Cullen}
\affiliation{Institute for Astronomy, University of Edinburgh, Royal Observatory, Edinburgh, EH9 3HJ, UK}
\email{fergus.cullen@ed.ac.uk}

\author[0000-0002-9551-0534]{Emma Curtis-Lake}
\affiliation{Centre for Astrophysics Research, Department of Physics, Astronomy and
Mathematics, University of Hertfordshire, Hatfield AL10 9AB, UK}
\email{e.curtis-lake@herts.ac.uk}

\author[0000-0003-2842-9434]{Romeel Dav\'e}
\affiliation{Institute for Astronomy, University of Edinburgh, Royal Observatory, Edinburgh, EH9 3HJ, UK}
\email{rdave@ed.ac.uk}

\author[0000-0002-3324-4824]{Rebecca L. Davies}
\affiliation{Centre for Astrophysics and Supercomputing, Swinburne University of Technology, John Street, Hawthorn, 3122, VIC, Australia}
\email{rdavies@swin.edu.au}

\author[0000-0003-0348-2917]{Miroslava Dessauges-Zavadsky}
\affiliation{Department of Astronomy, University of Geneva, Chemin Pegasi 51, 1290 Versoix, Switzerland}\email{}

\author[0000-0001-5414-5131]{Mark Dickinson}
\affiliation{NSF National Optical-Infrared Astronomy Research Laboratory, 950 N. Cherry Ave., Tucson, AZ 85719, USA}
\email{mark.dickinson@noirlab.edu}

\author[0000-0002-7622-0208]{Callum T. Donnan}
\affiliation{NSF National Optical-Infrared Astronomy Research Laboratory, 950 N. Cherry Ave., Tucson, AZ 85719, USA}
\email{callum.donnan@noirlab.edu}

\author[0009-0009-8105-4564]{Qiao Duan}
\affiliation{Kavli Institute for Cosmology, University of Cambridge, Madingley Road, Cambridge CB3 0HA, UK}
\email{qd231@cam.ac.uk}

\author[0000-0001-6889-8388]{Kenneth J. Duncan}
\affiliation{Institute for Astronomy, University of Edinburgh, Royal Observatory, Edinburgh, EH9 3HJ, UK}
\email{kdun@roe.ac.uk}

\author[0000-0002-1404-5950]{James S. Dunlop}
\affiliation{Institute for Astronomy, University of Edinburgh, Royal Observatory, Edinburgh, EH9 3HJ, UK}
\email{james.dunlop@ed.ac.uk}

\author[0000-0002-9382-9832]{Andreas L. Faisst}
\email{afaisst@caltech.edu}
\affiliation{IPAC, California Institute of Technology, 1200 E. California Blvd. Pasadena, CA 91125, USA}

\author[0000-0001-8519-1130]{Steven L. Finkelstein}
\affiliation{Department of Astronomy, University of Texas at Austin, 2515 Speedway, Austin, TX 78712, USA}
\affiliation{Cosmic Frontier Center, The University of Texas at Austin, Austin, TX 78712, USA}\email{}

\author[0000-0002-0159-2613]{Sophia Flury}
\affiliation{Institute for Astronomy, University of Edinburgh, Royal Observatory, Edinburgh, EH9 3HJ, UK}
\email{sflury@roe.ac.uk}

\author[0000-0002-9656-1800]{Anna R. Gallazzi}
\affiliation{INAF-Osservatorio Astrofisico di Arcetri, Largo Enrico Fermi 5, 50126, Firenze, Italy}
\email{anna.gallazzi@inaf.it}

\author[0000-0002-3351-1216]{Adriana Gargiulo}
\affiliation{INAF - Istituto di Astrofisica Spaziale e Fisica Cosmica,Via Alfonso Corti 12, 20133, Milano, Italy}
\email{adriana.gargiulo@inaf.it}

\author[0000-0001-9885-4589]{Steven Gillman}
\affiliation{Cosmic Dawn Center (DAWN), Copenhagen, Denmark}
\affiliation{DTU-Space, Elektrovej, Building 328, 2800, Kgs. Lyngby, Denmark}
\email{srigi@space.dtu.dk}

\author[0000-0002-3254-9044]{Karl Glazebrook}
\affiliation{JWST Australian Data Centre, Centre for Astrophysics and Supercomputing, Swinburne University of Technology, PO Box 218, Hawthorn, VIC 3122, Australia}
\email{kglazebrook@swin.edu.au}

\author[0000-0003-4242-8606]{Ansh R. Gupta}
 \affiliation{Department of Astronomy, University of Texas at Austin, 2515 Speedway, Austin, TX 78712, USA}
 \affiliation{Cosmic Frontier Center, The University of Texas at Austin, Austin, TX 78712, USA}
 \email[]{anshrg@utexas.edu}
 
\author[0000-0001-6145-5090]{Nimish P. Hathi}
\affiliation{U.S. National Science Foundation (NSF), Alexandria, VA 22314, USA}
\email{nphathi@gmail.com}

\author[0000-0002-9389-7413]{Kasper E. Heintz}
\affiliation{DTU Space, Technical University of Denmark, Elektrovej 327, DK2800 Kgs. Lyngby, Denmark}
\affiliation{Cosmic Dawn Center (DAWN), Copenhagen, Denmark}
\affiliation{Niels Bohr Institute, University of Copenhagen, Jagtvej 128, DK-2200, Copenhagen, Denmark}\email{}

\author[0009-0006-3003-9688]{Isaac J. B. Holst}
\affiliation{Institute for Astronomy, University of Edinburgh, Royal Observatory, Edinburgh, EH9 3HJ, UK}
\email{isaac.holst@ed.ac.uk}

\author[0000-0002-9453-0381]{Kei Ito}
\affiliation{Cosmic Dawn Center (DAWN), Copenhagen, Denmark}
\affiliation{DTU-Space, Elektrovej, Building 328, 2800, Kgs. Lyngby, Denmark}
\email{keiit@dtu.dk}

\author[0000-0002-8896-6496]{Christian Kragh Jespersen}
\affiliation{David A. Dunlap Department of Astronomy \& Astrophysics, University of Toronto, Toronto, ON M5S 3H4, Canada}
\affiliation{Department of Statistical Sciences, University of Toronto, Toronto, ON M5S 3G3, Canada}
\affiliation{Canadian Institute for Theoretical Astrophysics, University of Toronto, 60 St George Street, Toronto, M5S 3H8, Ontario, Canada}
\email{christian.jespersen@utoronto.ca}

\author[orcid=0000-0001-9187-3605,gname=Jeyhan, sname=Kartaltepe]{Jeyhan S. Kartaltepe}
\affiliation{Laboratory for Multiwavelength Astrophysics, School of Physics and Astronomy, Rochester Institute of Technology, 84 Lomb Memorial
Drive, Rochester, NY 14623, USA}
\email{jeyhan@astro.rit.edu}

\author[0000-0002-6610-2048]{Anton M. Koekemoer}
\affiliation{Space Telescope Science Institute, 3700 San Martin Drive, Baltimore, MD 21218, USA}
\email{koekemoer@stsci.edu}

\author[0000-0002-7613-9872]{Mariska Kriek}
\affiliation{Leiden Observatory, Leiden University, P.O. Box 9513, 2300 RA Leiden, The Netherlands}
\email{kriek@strw.leidenuniv.nl}

\author[0000-0001-5710-8395]{Danial Langeroodi}
\affiliation{Kavli Institute for Cosmology, University of Cambridge, Madingley Road, Cambridge CB3 0HA, UK}
\email{dr659@cam.ac.uk}

\author[0000-0001-6755-1315]{Joel Leja}
\affiliation{Department of Astronomy \& Astrophysics, The Pennsylvania State University, University Park, PA 16802, USA} 
\affiliation{Institute for Computational \& Data Sciences, The Pennsylvania State University, University Park, PA 16802, USA}
\affiliation{Institute for Gravitation and the Cosmos, The Pennsylvania State University, University Park, PA 16802, USA}\email{}

\author[0000-0002-2725-302X]{Yongming Liang}
\affiliation{National Astronomical Observatory of Japan, 2-21-1 Osawa, Mitaka, Tokyo 181-8588, Japan}
\email{yongming.liang@nao.ac.jp}

\author[0000-0003-0771-763X]{Yuankang Liu}
\affiliation{Institute for Computational Cosmology, Department of Physics, Durham University, South Road, Durham DH1 3LE, UK}
\email{yuankang.liu@durham.ac.uk}

\author[0000-0003-4368-3326]{Derek J. McLeod}
\affiliation{Institute for Astronomy, University of Edinburgh, Royal Observatory, Edinburgh, EH9 3HJ, UK}
\email{derek.mcleod@ed.ac.uk}

\author[0009-0005-9742-2318]{Ross McLure}
\affiliation{Institute for Astronomy, University of Edinburgh, Royal Observatory, Edinburgh, EH9 3HJ, UK}
\email{rmclure@ed.ac.uk}

\author[0000-0002-0984-7713]{Yurina Nakazato}
\affiliation{Center for Computational Astrophysics, Flatiron Institute, 162 5th Avenue, New York, NY 10010, USA}
\email{ynakazato@flatitoninstitute.org}

\author[0000-0003-2804-0648]{Themiya Nanayakkara}
\affiliation{Sydney Institute for Astronomy, School of Physics, The University of Sydney, Sydney, New South Wales, Australia}\email{}

\author[]{Ga\"el Noirot}
\affiliation{Space Telescope Science Institute, 3700 San Martin Drive, Baltimore, MD 21218, USA}
\email{gnoirot@stsci.edu}

\author[0000-0003-2000-3420]{Katherine Ormerod}
\affiliation{Astrophysics Research Institute, Liverpool John Moores University, 146 Brownlow Hill, Liverpool, L3 5RF, UK}
\email{K.Ormerod@2023.ljmu.ac.uk}

\author[0000-0002-9651-5716]{Richard Pan}
\affiliation{Department of Physics and Astronomy, Tufts University, 574 Boston Ave., Medford, MA 02155, USA}\email{}

\author[0000-0002-8435-9402]{Minjung Park}
\affiliation{Kavli Institute for Cosmology, University of Cambridge, Madingley Road, Cambridge CB3 0HA, UK}
\affiliation{Cavendish Laboratory, University of Cambridge, 19 JJ Thomson Avenue, Cambridge CB3 0HE, UK}
\email{mp2246@cam.ac.uk}

\author[orcid=0000-0002-1283-8420]{Robert Pascalau}
\affiliation{Kavli Institute for Cosmology, University of Cambridge, Madingley Road, Cambridge CB3 0HA, UK}
\email{rgp34@cam.ac.uk}

\author[0000-0003-4528-5639]{Pablo G. P\'erez-Gonz\'alez}
\affiliation{Centro de Astrobiolog\'{\i}a (CAB), CSIC-INTA, Ctra. de Ajalvir km 4, Torrej\'on de Ardoz, E-28850, Madrid, Spain}
\email{pgperez@cab.inta.csic.es}

\author{Eniko Regos} 
\affiliation{Konkoly Observatory, HUN-REN, Budapest, Konkoly Thege M. 15-17, H-1121}
\email{regos@konkoly.hu}

\author[0000-0002-5104-8245]{Pierluigi Rinaldi}
\email{prinaldi@utexas.edu}
\affiliation{Department of Astronomy, University of Texas at Austin, 2515 Speedway, Austin, TX 78712, USA}
\affiliation{Cosmic Frontier Center, The University of Texas at Austin, Austin, TX 78712, USA}

\author[0009-0009-2671-4160]{Lucie E. Rowland}
\email{lucie.rowland@eso.org}
\affiliation{European Southern Observatory (ESO), Karl-Schwarzschild Str. 2, 85748 Garching, Germany}

\author[0000-0001-7883-8434]{Kate Rowlands}
\affiliation{AURA for ESA, Space Telescope Science Institute, 3700 San Martin Drive, Baltimore, MD 21218, USA}
\affiliation{William H. Miller III Department of Physics and Astronomy, Johns Hopkins University, Baltimore, MD 21218, USA}
\email{krowlands@stsci.edu}

\author[0000-0003-4792-9119]{Ryan L. Sanders}
\affiliation{Department of Physics and Astronomy, University of Kentucky, 505 Rose Street, Lexington, KY 40506, USA}\email{}

\author[0000-0003-2227-4902]{Marc Sarzi}
\affiliation{Armagh Observatory and Planetarium, College Hill, Armagh, BT61 9DG, N. Ireland, UK}
\email{marc.sarzi@armagh.ac.uk}

\author[0000-0003-3509-4855]{Alice E. Shapley}
\affiliation{University of California, Los Angeles, 475 Portola Plaza, Los Angeles, CA, 90095, USA}
\email{aes@astro.ucla.edu}

\author[0009-0004-0844-0657]{Maya Skarbinski}
\affiliation{William H. Miller III Department of Physics and Astronomy, Johns Hopkins University, Baltimore, MD 21218, USA}
\email{mskarbi1@jh.edu}

\author[orcid=0000-0001-7540-1544]{Martje Slob}
\affiliation{Leiden Observatory, Leiden University, P.O. Box 9513, 2300 RA Leiden, The Netherlands}
\email{slob@strw.leidenuniv.nl}

\author[0000-0002-0827-9769]{Thomas M. Stanton}
\affiliation{Institute for Astronomy, University of Edinburgh, Royal Observatory, Edinburgh, EH9 3HJ, UK}
\email{t.stanton@ed.ac.uk}

\author[0000-0001-6561-9443]{Yang Sun}
\affiliation{Steward Observatory, University of Arizona, 933 North Cherry Avenue, Tucson, AZ 85719, USA}
\email{sunyang@arizona.edu}

\author[0009-0005-4175-4890]{Chamilla Terp}
\affiliation{DTU Space, Technical University of Denmark, Elektrovej 327, DK2800 Kgs. Lyngby, Denmark}\email{}

\author[0000-0001-6477-4011]{Francesco Valentino}
\affiliation{Cosmic Dawn Center (DAWN), Copenhagen, Denmark}
\affiliation{DTU-Space, Elektrovej, Building 328, 2800, Kgs. Lyngby, Denmark}
\email{fmava@dtu.dk}

\author[0000-0002-5027-0135]{Arjen van der Wel}
\affiliation{Department of Physics and Astronomy, Universiteit Gent, Proeftuistraat 86, 9000 Gent, Belgium}
\email{arjen.vanderwel@ugent.be}

\author[0000-0002-8445-4397]{Jorick S. Vink}
\affiliation{Armagh Observatory and Planetarium, College Hill, Armagh, BT61 9DG, N. Ireland, UK}
\email{jorick.vink@armagh.ac.uk}

\author[0000-0001-9269-5046]{Bingjie Wang}
\affiliation{Department of Astrophysical Sciences, Princeton University, Princeton, NJ 08544, USA}
\email{bjwang@princeton.edu}

\author{Vivienne Wild}
\affiliation{School of Physics \& Astronomy, University of St Andrews, St Andrews, KY16 9SS, UK}
\email{vw8@st-andrews.ac.uk}

\author[0000-0003-3903-6935]{Stephen M. Wilkins}
\email{S.Wilkins@sussex.ac.uk}
\affiliation{Astronomy Centre, University of Sussex, Falmer, Brighton BN1 9QH, UK}

\author{Po-Feng Wu}
\affiliation{Graduate Institute of Astrophysics, National Taiwan University, Taipei 10617, Taiwan}
\email{wupofeng@phys.ntu.edu.tw}

\author[0000-0003-1734-8356]{Stefano Zibetti}
\affiliation{INAF-Osservatorio Astrofisico di Arcetri, Largo Enrico Fermi 5, 50126, Firenze, Italy}
\email{stefano.zibetti@inaf.it}

\begin{abstract}
\noindent We argue for a significant investment of JWST time into an extremely deep medium-resolution NIRSpec MSA survey targeting a large, representative sample of high-redshift $(z>1.5)$ galaxies, specifically intended to obtain high-SNR continuum detections. This is only possible with JWST, and no planned future instrumentation has similar capabilities, meaning that obtaining such observations with JWST must be urgently prioritised. The vast majority of existing archival NIRSpec observations are not deep enough for continuum science, being focused on the detection of emission lines. Deeper continuum data opens up the ability to constrain a much wider range of key galaxy physical parameters, including stellar population ages and metallicities, stellar kinematics and neutral gas physical properties, inflows and outflows. Such measurements will provide crucial insight into quenching physics, galaxy dynamical masses, merger histories, pristine early star-formation, reionization and the physics of Little Red Dots, as well as enabling all of the emission-line science that can be achieved with shallower observations. To facilitate future survey planning, we present an empirically derived relationship between rest-optical continuum SNR in JWST NIRSpec medium-resolution data, rest-optical apparent magnitude and exposure time, which we show to be independent of redshift (and hence the grating used). We then provide a 1000-hour sketch survey plan based on achieving the continuum SNR $=20$/pixel necessary to constrain key physical properties and scaling relations for a large galaxy sample at $z>1.5$. We finally discuss NIRSpec sky subtraction strategies, which are of critical importance for continuum SNR.

\newpage
\end{abstract}


\section{\textbf{The unique nature of JWST spectroscopy}}

The spectroscopic capabilities onboard JWST are the area in which it most thoroughly outclasses all previous instrumentation. Ground-based spectroscopy in the near infrared (NIR) is plagued by atmospheric emission and absorption processes, seriously limiting the depth and wavelength coverage that can be achieved. This becomes progressively worse with increasing wavelength, and observations at $\lambda>2.5\,\mu$m in particular are all but impossible from the ground. At these wavelengths, JWST is more sensitive by a factor of $>1000$. This has revolutionised the study of high-redshift galaxies, for which the crucial rest-frame optical spectral features typically used to infer physical properties are redshifted into the NIR (in particular at $z>1.5$; see Section \ref{sec:technical}).

The spectroscopic capabilities of JWST are a precious and worryingly finite resource, given the limited lifetime of the telescope. The next proposed future flagship mission, Habitable Worlds Observatory (HWO), is more UV-optical focused, and is unlikely to provide NIR spectroscopic capabilities to rival JWST, with no capability at all planned for $\lambda>2.5\,\mu$m. Based on the length of the JWST development cycle and the planned launch of HWO in the 2040s, \textcolor{blue}{\textbf{it is likely that the NIR spectroscopic capabilities of JWST will not be equalled or surpassed for at least the next 50 years. It is therefore essential that we now prioritise building a comprehensive reference sample of ultra-deep spectroscopic observations for $\mathbf{z>1.5}$ galaxies.}}

\section{\textbf{The need for medium resolution}}

Inferring galaxy physical properties from observational data is a complex and challenging exercise. The degree of success with which this can be carried out is a strong function of both the signal-to-noise ratio (SNR) and the spectral resolving power ($R=\lambda/\Delta\lambda)$ of the data that can be obtained. Imaging data, (obtained for example via the NIRCam wide and medium filters) can be used to infer basic properties, such as photometric redshifts, stellar masses, star-formation rates (SFRs) and rest-frame magnitudes (though even these are far more reliably inferred from spectroscopy). Photometric data however cannot be used to measure subtler galaxy physical properties, such as gas-phase and stellar metallicities, star-formation histories (SFHs), kinematics, gas-phase ionization states, gas inflow/outflow properties, or levels of dust attenuation. The most important reason for this is the age-metallicity-dust degeneracy (e.g., \citealt{Conroy2013}): all three of these variables have similar effects on galaxy spectral shapes, meaning they cannot be reliably separated with photometric data alone. This is because photometric data cannot resolve the individual emission and absorption features in galaxy spectra necessary to break this degeneracy, measurements of which are also needed to infer all of the other properties listed above.

\begin{figure}[t]
\begin{center}
\includegraphics[width=\textwidth]{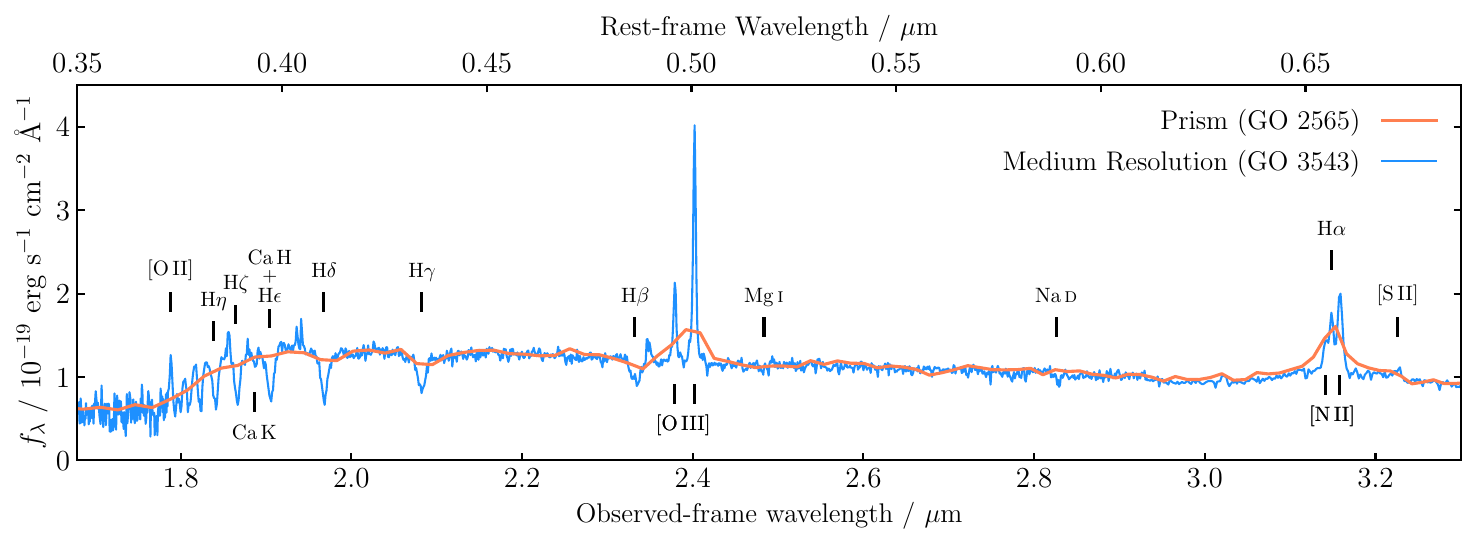}
\end{center}
\vspace{-0.5cm}
\caption{A direct comparison of NIRSpec prism and medium-resolution grating data for a galaxy at $z=3.797$. Key emission and absorption features necessary to derive physical properties are labelled. It can be seen that the $R\simeq30-300$ prism data fails to resolve both the Balmer series lines in the 4000\AA\, break region, crucial for age measurement, as well as the [O\,\textsc{iii}]+H$\beta$ and [N\,\textsc{ii}]+H$\alpha$ emission line complexes crucial for SFR, gas-phase metallicity, ionization state and kinematic measurements.}\label{fig:prism_vs_mr}
\end{figure}

A particular innovation of NIRSpec is its prism mode, which provides low-resolution spectroscopy ($R\simeq30-300$) over the instrument's full $\lambda\simeq1-5\,\mu$m range. This is valuable for confirming uncertain photometric redshifts and studying extremely faint ultra-high-redshift galaxies. The low resolution of the prism mode is however still insufficient for resolving many of the crucial spectral features necessary for measuring subtle galaxy physical properties, which requires $R\gtrsim1000$ (e.g., \citealt{Pacifici2012}). An example of this is presented in Fig. \ref{fig:prism_vs_mr}, which shows a direct comparison of prism-mode data with medium-resolution grating spectroscopy ($R=1000$) for the same galaxy at $z=3.717$. It can be seen that prism data fails to resolve the individual higher-order Balmer absorption features in the Balmer/4000\,\AA\, break region, which are crucial for the measurement of galaxy SFHs. It also fails to resolve the individual emission lines in the [O\,\textsc{iii}]+H$\beta$ and [N\,\textsc{ii}]+H$\alpha$ complexes, which is necessary to measure the dust and metal contents and ionization state of gas in galaxies. For this reason, it must be emphasised as strenuously as possible that \textcolor{blue}{\textbf{NIRSpec prism data is not a suitable replacement for medium-resolution grating data in most cases.}}

\section{\textbf{The scientific case for continuum observations}}\label{sec:science}

The bulk of existing NIRSpec medium-resolution observations are designed around detection of emission lines in order to study gas-phase metallicities, ionization states and kinematics, as well as AGN properties. Detecting continuum emission from galaxies at high SNR, and hence absorption lines as well as emission lines, typically requires substantially longer integrations. The payoff for this extra investment is threefold, enabling investigations into:
\textbf{\textcolor{blue}{
\begin{enumerate}
\setlength\itemsep{0em}
\item The star formation and chemical enrichment histories of galaxy stellar populations
\item Stellar kinematics, halo masses and merger histories
\item The presence and properties of neutral gas in and around galaxies
\end{enumerate}
}}

\noindent Furthermore, observations deep enough to detect continuum emission at high SNR are guaranteed to detect all emission lines present, even where very weak, enabling all of the investigations into gas-phase properties listed above.

\subsection{\textbf{Star-formation and chemical enrichment histories}}

Continuum emission from stars, and the superimposed absorption lines from stellar atmospheres, provide a wealth of information about the stellar populations in galaxies. The most important quantities that can be constrained are the SFH and chemical enrichment history, often discussed in terms of their average values: the stellar age and stellar metallicity. Galaxy SFHs can be used to understand when the bulk of the stellar mass in a galaxy was formed, and to compare current levels of star formation with historical ones. In particular, the archaeological downsizing trend (stellar mass vs stellar age relationship) is a key test for the quenching physics employed by models and simulations of galaxy formation (e.g., \citealt{gallazzi_ages_2005, nelson_first_2018,carnall_vandels_2019,leung_jwst_2026}). More detailed comparisons of this nature between simulations and observations are now crucial, as a wide variety of models employing different physical recipes can now well reproduce the (more straightforward) galaxy stellar mass function (e.g., \citealt{somerville_physical_2015}).

Chemical enrichment histories can be extracted by measuring the stellar metallicities of different-age populations, tracing the full history of the baryon cycle in galaxies, in contrast with the gas-phase metallicity, which provides a snapshot at the time of observation. \textcolor{blue}{\textbf{These two quantities in combination can constrain a wide variety of key processes regulating the baryon cycle, such as gas inflow rates and stellar and AGN feedback}}, as well as providing constraints on star-formation timescales when the abundances of multiple elements are considered together (e.g., the [$\alpha$/Fe] ratio; \citealt{thomas_environment_2010,beverage_elemental_2021,shapley_aurora_2025}).

Before JWST, measurements of galaxy stellar ages and metallicities beyond $z=1$ were extremely rare, as observations deep enough to measure continuum emission were impractical to obtain, except for the very brightest galaxies (e.g., \citealt{kriek_massive_2016,valentino_quiescent_2020}). More often than not, stacking together large numbers of sources was necessary to detect continuum emission (e.g., \citealt{schreiber_near_2018, cullen_vandels_2019, carnall_stellar_2022}). Driven by JWST, progress over the past 5 years has been rapid, with large numbers of studies beginning to constrain how stellar age and metallicity evolve with stellar mass during the first few billion years after the Big Bang (e.g., \citealt{park_widespread_2024,slob_jwst-suspense_2024,beverage_carbon_2025,hamadouche_deepdive_2026,leung_jwst_2026}) and even the beginnings of spatially resolved work (e.g., \citealt{cheng_clues_2026,edward_magaz3ne_2026}).

\subsection{\textbf{Stellar kinematics, halo masses and merger histories}}

Absorption-line spectroscopy can also be used to measure stellar kinematics, constraining in particular the dynamical masses and merger histories of galaxies. The majority of such work to date with JWST has focused on early quiescent galaxies, which have largely been found to be fast rotating, suggesting a past free of significant merger activity (e.g., \citealt{slob_fast_2025,pascalau_when_2026}). However, recent work by \cite{forrest_massive_2026} has demonstrated this is not always the case, with the discovery of a slow rotating ultra-massive quiescent galaxy at $z\simeq3.5$. \textcolor{blue}{\textbf{This demonstrates the inadequate sizes of current samples for gaining a true picture of early massive galaxy evolution, highlighting again the need for a much larger and more systematic ultra-deep medium-resolution observing campaign.}}

\subsection{\textbf{The presence and properties of neutral gas in and around galaxies}}

Neutral gas is the main fuel reservoir in galaxies, condensing to form molecular gas that then collapses into stars. Often studied in the local Universe both in the radio regime via CO emission and in the UV-optical via interstellar medium (ISM) absorption lines, extending such observations to $z>1$ is more challenging, with progress made in the radio regime (e.g., \citealt{carilli_cool_2013}), but virtually none in the UV-optical, owing to a lack of continuum observations. JWST has again revolutionised this situation, providing ISM absorption line detections of neutral gas in hundreds of high-redshift galaxies (e.g., \citealt{sun_census_2026,sapori_outflows_2026}). 

The kinematics of such features can be used to study neutral gas inflows and outflows, with widespread interest developing into whether strong AGN-driven outflows are responsible for quenching the abundance of $2 < z < 7$ quiescent galaxies discovered by JWST (e.g., \citealt{belli_star_2024,davies_jwst_2024,wu_ejective_2025,valentino_gas_2025,taylor_jwst_2026,davies_galactic_2026,zhu_there_2026}). Powerful outflows are also being detected by JWST via absorption lines in rest-frame UV spectroscopy of lower-mass and very highly star forming galaxies within the first billion years of cosmic history (e.g., \citealt{chen_spurs1_2026,chen_spurs_2026,keerthi_vasan_g_jwst_2026}). 

Medium-resolution spectroscopy is also the clear next step for the study of nebular continuum emission (e.g., \citealt{cameron_nebular_2024,katz_21_2025}), which is challenging to reliably separate from the stellar and emission-line components around the Balmer break/jump region using NIRSpec prism data alone. \textcolor{blue}{\textbf{Such observations are crucial for building our understanding of the first stages of primordial and potentially population-\textsc{iii} star formation in the early Universe}}, in particular providing further insight into the potential role of low-metallicity very massive stars (e.g., \citealt{vink_very_2025,cullen_jwst_2025,marques-chaves_signatures_2026}).

Another area in which neutral gas absorption line studies have proven crucial is in unravelling the mystery of little red dots (LRDs). Such objects commonly display broad Balmer and metal absorption lines, thought to result from high-density cool gas in proximity to the central engine (e.g., \citealt{matthee_little_2024,kocevski_rise_2025, wang_balmer_2026,deugenio_blackthunder_2026}). Detailed absorption line profiles for LRDs, only available at medium resolution, are crucial for constraining the physical properties of their gas (e.g., \citealt{sneppen_inside_2026,tang_spurs_2026}). The extremely red colours of LRDs and the fact they are almost exclusively found at very high redshift mean that all further study without the $\lambda>2.5\,\mu$m spectroscopic capabilities of JWST will be impossible.

A final key area is the study of cosmic reionization. Observations of the Lyman alpha damping wing can be used to measure the ionization state of gas in the intergalactic medium during reionization, as well as the extent of ionized bubbles in the immediate vicinity of galaxies (e.g., \citealt{umeda_jwst_2024,heintz_strong_2024,keating_jwst_2024,keating_origin_2024}). The improved resolution over prism data assists with breaking the multi-parameter degeneracies intrinsic to this work (including by providing more detailed shape information for the Lyman alpha line where present), providing stronger constraints on key derived quantities (e.g., \citealt{witstok_witnessing_2025,witstok_oasis_2026}).

\section{\textbf{The need for a multi-cycle treasury programme}}
Virtually all of the recent JWST work discussed in Section \ref{sec:science} is based on small samples (from individual galaxies to $\simeq10$s of objects at most) from dedicated programmes, with the bulk of archival data not deep enough to detect continuum emission at high SNR. \textcolor{blue}{\textbf{The small sizes of available samples precludes truly statistical constraints on key scaling relations with stellar mass (e.g., for stellar ages, metallicities, kinematics, outflow incidences and velocities). Furthermore, uniform, clean and well-understood selection functions are crucial for such work, in order to gain a representative understanding of the galaxy population.}}

For example, such scaling relations typically exhibit intrinsic scatter at levels of $\simeq0.1-0.3$ dex (e.g., \citealt{schreiber_star-forming_2020,hamadouche_connection_2023}). Taking a representative scatter of 0.2 dex, approximately 25 objects per bin in stellar mass and redshift would be required to achieve average values robust to a $\simeq10$ per cent level (0.04 dex). Good constraints on such scaling relations might therefore be achieved with $\simeq100$ well-chosen galaxies per redshift bin, based on e.g., $4\,\times\,0.3-$dex stellar mass bins from $10.3 < \mathrm{log}_{10}(M_*/\mathrm{M_\odot}) < 11.5$. Ideally, considerably more objects with deep continuum spectroscopy would be available, in order to begin constraining subtler co-dependencies between key parameters that each individually vary strongly with stellar mass. For example, multiple $\simeq25-$object bins of galaxy size within a fixed redshift and stellar mass interval would allow investigation into how size correlates with properties such as stellar metallicity, providing deeper insight into the underlying physical processes.

\vspace{0.2cm}

\noindent \begin{minipage}{0.46\textwidth}
\vspace{-0.3cm}

\includegraphics[width=\textwidth]{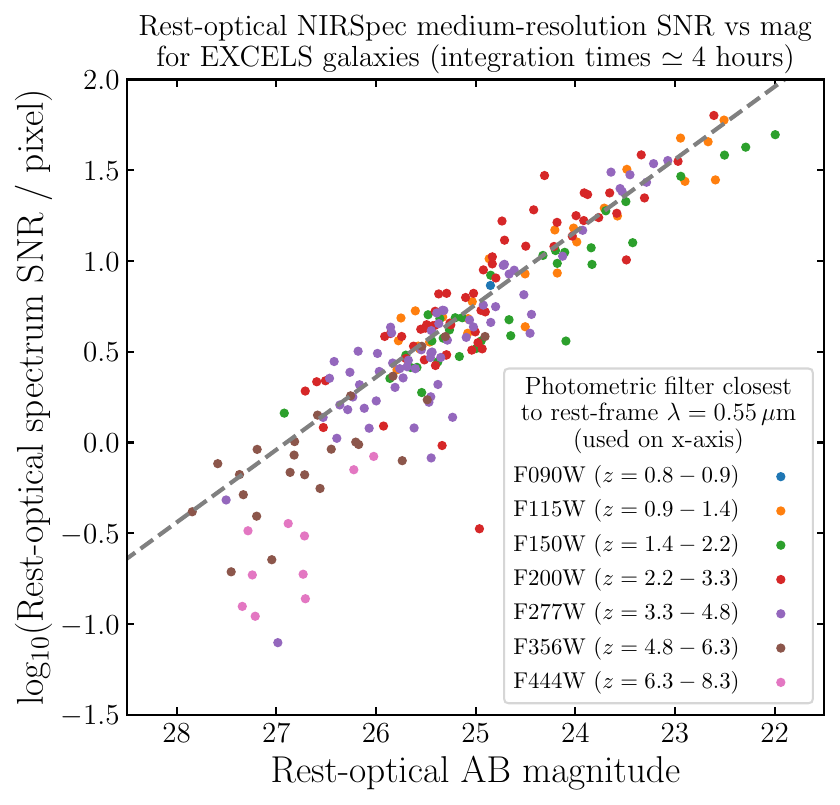}
\captionof{figure}{NIRSpec medium resolution SNR per pixel in the rest-frame optical $(0.51 < \lambda_\mathrm{rest} < 0.64\,\mu$m) as a function of apparent magnitude (taken as the closest NIRCam broad band to $\lambda_\mathrm{rest}=0.55\,\mu$m) for EXCELS survey galaxies. The form of grey line of best fit is given in Equation \ref{eqn:snr}. \label{fig:snr_vs_mag}}

\end{minipage}
\hspace{0.02\textwidth}
\begin{minipage}{0.51\textwidth}
\vspace{-0.3cm}

\includegraphics[width=\textwidth]{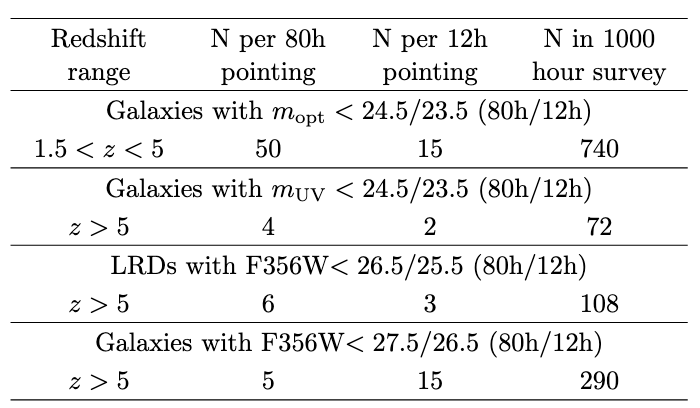}

\captionof{table}{The number of objects to be observed via the sketch survey strategy outlined in Section \ref{sec:technical}. Objects in the top 2 classes are selected such that we will obtain SNR $\geq$ 20 per pixel at $R=1000$ (assuming 2 pixels per resolution element) in a 20 hour integration. Each 80 hour pointing would obtain 20 hour exposures with the G140M, G235M and G395M gratings. Each 12 hour pointing would obtain 3 hour exposures for objects $\geq1$ mag brighter, reaching the same minimum SNR. Note that the 12 hour pointing masks are only $\simeq60$ per cent filled by these brighter target classes, and would hence also have room for fainter fillers. Apparent magnitudes $m_\mathrm{opt}$ and $m_\mathrm{UV}$ are defined using the closest NIRCam broad band to $\lambda_\mathrm{rest}=0.55\,\mu$m and $0.15\,\mu$m respectively. }

\label{table:ndensity}

\end{minipage}

\vspace{0.1cm}

To further illustrate this point, \cite{ito_deepdive_2026} recently conducted an archival search for NIRSpec medium-resolution spectroscopic observations of quiescent galaxies at $1 < z < 5$, finding 136 objects with such publicly available data. This level of archival data availability is inadequate for two key reasons. Firstly, significant evolution is expected in key scaling relations during the $\simeq5$ Gyr between $z=5$ and $z=1$, which cannot be captured unless sufficient spectra are available to define multiple redshift bins in this interval. Secondly, the vast majority of such extant data does not reach high enough continuum SNR to conduct the full range of scientific analyses discussed in Section \ref{sec:science}. \textcolor{blue}{\textbf{It is therefore crucial that a large, systematically selected high-redshift galaxy sample is surveyed to continuum depth at medium resolution during the remaining lifetime of JWST.}} As will be demonstrated in Sections \ref{sec:snr} and \ref{sec:technical} below, such a reference sample will require a significantly greater time investment than would be possible via the ordinary GO process, but would naturally also serve very well a variety of other science cases that typically require shorter integrations (e.g., emission line studies), including for extremely faint sources.




\section{\textbf{Required SNR and integration times}}\label{sec:snr}

Robust constraints on continuum-derived properties (e.g., $\simeq0.1$ dex precision measurements of stellar ages and metallicities) require rest-optical continuum SNR $\gtrsim20$ per pixel (assuming 2 pixels per resolution element) at $R=1000$ (e.g., \citealt{Pacifici2012,leung_jwst_2026}). 
To quantify the integration times required to reach such rest-optical SNRs with the NIRSpec MSA, we conduct an empirical analysis using data from the EXCELS survey \citep{carnall_jwst_2024} as this spans the whole $1-5\,\mu$m observed wavelength range of NIRSpec at medium resolution, as well as containing galaxies with a wide range of magnitudes, redshifts and spectral types.

We calculate the rest-frame optical SNR per pixel from EXCELS spectra as the mean value across $\lambda_\mathrm{rest}=0.51-0.64\,\mu$m, retaining only galaxies that have $>50$ per cent coverage of this wavelength range. We show these values in Fig. \ref{fig:snr_vs_mag} as a function of apparent magnitude in the NIRCam broad band closest to rest-frame $\lambda=0.55\,\mu$m for each object, $m_\mathrm{opt}$. EXCELS observed for $\simeq4$ hours with all of G140M, G235M and G395M (The G235M integration time was $\simeq5.5$ hours, however this should only result in SNRs increased by $\lesssim0.1$ dex). Under these conditions it can be seen that SNR per pixel is a simple linear function of $m_\mathrm{opt}$, irrespective of observed redshift (and hence the grating into which the rest-optical wavelength range falls). We thus derive the following relationship between rest-optical continuum SNR, NIRSpec medium resolution exposure time, $t_\mathrm{exp}$, and $m_\mathrm{opt}$, shown in gray in Fig. \ref{fig:snr_vs_mag}:

\begin{equation}
\mathrm{log_{10}\bigg(\dfrac{SNR}{pixel}}\bigg) = 10.45 + 0.5\times\mathrm{log}_{10}\bigg(\dfrac{t_\mathrm{exp}}{\mathrm{hours}}\bigg)- 0.4\times m_\mathrm{opt}\label{eqn:snr}
\end{equation}\noindent This relationship can be combined with mass-completeness calculations to understand the integration times required to obtain at least a given SNR for a complete sample down to a given stellar mass limit. To build a representative example, we follow the approach of \cite{stevenson_primer_2026}, finding that a magnitude limit of $m_\mathrm{opt}=24.5$ (requiring 20 hours per medium grating to reach SNR $=20$/pixel) yields 90 per cent mass-completeness limits of log$_{10}(M_*/\mathrm{M}_\odot)\simeq10.3$ across $1.5 < z < 5$. At higher redshift massive galaxies are rarer, with focus shifting to LRDs and the rest-UV continua of highly star forming objects, which we also incorporate into our sketch survey plan below.

\section{\textbf{Survey planning considerations for continuum observations}}\label{sec:technical}

We here provide a brief demonstrative sketch of a large NIRSpec MSA medium-resolution continuum survey, aiming to achieve rest-optical SNR $\geq$ 20 per pixel for a large, representative sample. We propose a series of pointings including 20 hour integrations with all three medium-resolution gratings (for a total per-pointing execution time of $\simeq80$ hours assuming $\simeq30$ per cent overhead), with objects assigned to gratings using the following scheme:

\begin{minipage}{\textwidth}
\begin{center}
\includegraphics[width=\textwidth]{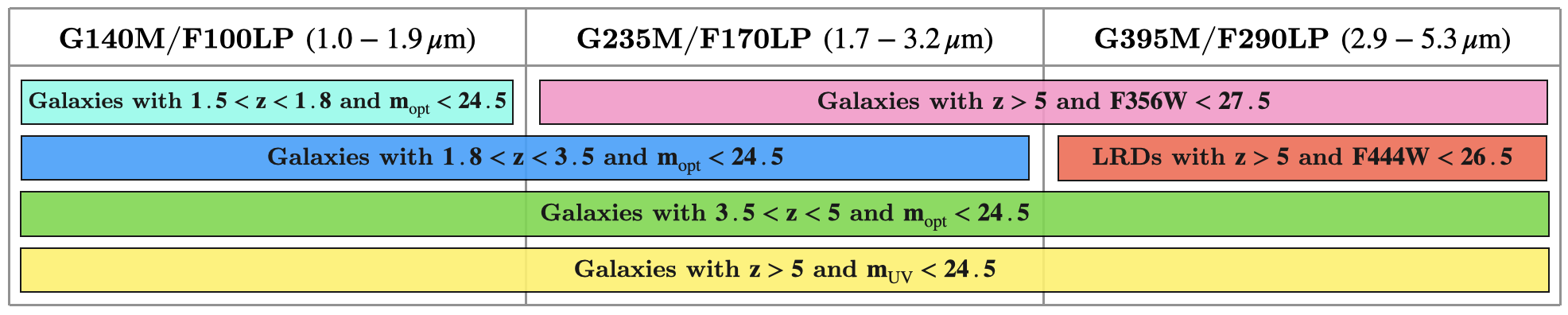}
\end{center}
\end{minipage}





\noindent We further propose a larger number of shorter pointings with 3 hour integrations in each grating (for a per-pointing execution time of $\simeq12$ hours) and the magnitude limits of each sample reduced by 1, maintaining the same SNR for a brighter sub-sample in order to produce improved sample statistics at the bright end.

The lower redshift limit of $z=1.5$ is chosen for two main reasons. Firstly, this is the lowest redshift at which the full key rest-optical wavelength range from the Balmer break at 3645\AA\ to [S\,\textsc{ii}] at 6725\AA\ is visible with our chosen grating/filter combinations. Secondly, below this redshift rest-optical continuum observations can be obtained from the ground, in particular with current and upcoming multi-object spectrographs (e.g., MOONS, PFS, WEAVE, 4MOST). 

Such a survey could be carried out in any combination of CANDELS fields, covered with NIRCam by JADES, CEERS and PRIMER. Using PRIMER UDS and COSMOS data we estimate sky densities for our two top-priority sub-samples: galaxies with $1.5 < z <5$ and $m_\mathrm{opt}<24.5$ have a sky density of $\simeq8.5$ per sq. arcmin, whereas galaxies with $z > 5$ and $m_\mathrm{UV}<24.5$ have a sky density of $\simeq0.2$ per sq. arcmin ($m_\mathrm{UV}$ is apparent magnitude in the NIRCam broad band closest to $\lambda_\mathrm{rest}=0.15\,\mu$m). The MSA has a $\simeq9.5$ sq. arcmin field of view, and can accommodate $\simeq60$ galaxies in medium-resolution mode (but see Section \ref{skysub}). We therefore confirm the MSA can be efficiently filled using this strategy. The exact numbers of objects observed from each target class can be optimised, e.g., by selecting sky areas that contain above average numbers of bright $z>5$ galaxies. We provide a rough breakdown of expected target numbers in Table \ref{table:ndensity} for a $10\times80$ hour plus $16\times12$ hour pointing strategy, totalling $\simeq1000$ hours execution time.


\subsection{\textbf{Sky subtraction considerations}}\label{skysub}
\enlargethispage{10\baselineskip}

A final, crucial consideration for continuum observations is the sky subtraction strategy, which is of primary importance for continuum SNR. Slitless spectroscopy, or other strategies that involve overlap between spectra from different objects, are highly inadvisable for continuum observations, as overlapping continuum is much harder to separate than overlapping series of emission lines. NIRSpec MSA observations typically use either 3 or 5 shutter slitlets, with sources nodded between each shutter facilitating ordinary `A$-$B' background subtraction. This increases background noise by a factor $\sqrt{1+1/(N-1)}$, where N is the number of nods (the EXCELS results shown in Fig. \ref{fig:snr_vs_mag} use $N=3$). An alternative possibility is `master' background subtraction, in which (usually) only single shutters are opened for each object, as well as a series of individual background shutters, which are averaged to create a master background with very little noise (e.g., \citealt{glidic_pixel--pixel_2025,belli_blue_2026}). This avoids the increase in random background noise that occurs in the A$-$B case, as well as roughly doubling the number of objects that can be observed at once using the MSA. The cost of this is the introduction of systematic uncertainties, which can be challenging to quantify and do not necessarily scale down with increasing exposure time, making the utility of this approach for very deep observations unclear. Master background subtraction is therefore a promising option, however detailed tests with similarly deep $\simeq20$ hour datasets would be highly advisable before committing large amounts of observing time to this strategy.

\newpage

\bibliography{sample701}{}
\bibliographystyle{aasjournalv7}

\end{document}